\documentclass[%
 reprint,
superscriptaddress,
 amsmath,amssymb,
 aps,
 prl,
]{revtex4-2}
\usepackage{array}
\usepackage{xr}
\usepackage[T1]{fontenc}
\usepackage{tgtermes}
\usepackage{graphicx}
\usepackage{dcolumn}
\usepackage{bm}
\usepackage{braket}
\usepackage[table]{xcolor}

\begin{document}

\newcommand{\tm}{$T_{\mathrm{m}}$\ }
\newcommand{\maintitle}{Assessment of First-Principles Methods in Modeling the Melting Properties of Water}
\title{\maintitle}
\author{Yifan Li}
\email{yifanl0716@gmail.com}
\affiliation{ 
Department of Chemistry, Princeton University, Princeton, NJ 08544, USA
}%
\author{Bingjia Yang}
\affiliation{ 
Department of Chemistry, Princeton University, Princeton, NJ 08544, USA
}%
\author{Chunyi Zhang}
\affiliation{ 
Eastern Institute of Technology, Ningbo, Zhejiang 315200, China
}%
\author{Axel Gomez}
\affiliation{ 
Department of Chemistry, Princeton University, Princeton, NJ 08544, USA
}%
\author{Pinchen Xie}
\affiliation{ 
Program in Applied and Computational Mathematics, Princeton University, Princeton, NJ 08544, USA
}
\author{Yixiao Chen}
\affiliation{ 
Program in Applied and Computational Mathematics, Princeton University, Princeton, NJ 08544, USA
}
\author{Pablo M. Piaggi}
\affiliation{CIC nanoGUNE BRTA, Tolosa Hiribidea 76, 20018 Donostia-San Sebastián, Spain}
\affiliation{Ikerbasque, Basque Foundation for Science, 48013 Bilbao, Spain}

\author{Roberto Car}
\email{rcar@princeton.edu}
\affiliation{ 
Department of Chemistry, Princeton University, Princeton, NJ 08544, USA
}%
\affiliation{ 
Program in Applied and Computational Mathematics, Princeton University, Princeton, NJ 08544, USA
}
\affiliation{ 
Department of Physics, Princeton University, Princeton, NJ 08544, USA
}%
\affiliation{ 
Princeton Institute for the Science and Technology of Materials, Princeton University, Princeton, NJ 08544, USA
}%

\date{\today}

\begin{abstract}
First-principles simulations have played a crucial role in deepening our understanding of the thermodynamic properties of water, and machine learning potentials (MLPs) trained on these first-principles data widen the range of accessible properties.
However, the capabilities of different first-principles methods are not yet fully understood due to the lack of systematic benchmarks, the underestimation of the uncertainties introduced by MLPs, and the neglect of nuclear quantum effects (NQEs). Here, we systematically assess first-principles methods by calculating key melting properties using path integral molecular dynamics (PIMD) driven by Deep Potential (DP) models trained on data from density functional theory (DFT) with SCAN, revPBE0-D3, SCAN0 and revPBE-D3 functionals, as well as from the MB-pol potential. We find that MB-pol is in qualitatively good agreement with the experiment in all properties tested, whereas the four DFT functionals incorrectly predict that NQEs increase the melting temperature. SCAN and SCAN0 slightly underestimate the density change between water and ice upon melting, but revPBE-D3 and revPBE0-D3 severely underestimate it. Moreover, SCAN and SCAN0 correctly predict that the maximum liquid density occurs at a temperature higher than the melting point, while revPBE-D3 and revPBE0-D3 predict the opposite behavior. Our results highlight limitations in widely used first-principles methods and call for a reassessment of their predictive power in aqueous systems. 

\end{abstract}

\maketitle

Water is arguably the most important substance on Earth.
Its rich phase diagram and anomalous properties continue to be the subject of intense research efforts~\cite{gallo_water_2016}. In the last decade, simulations based on first-principles electronic structure theory have been used with success in predicting and understanding the thermodynamic properties of water~\cite{ruiz_pestana_quest_2018, sun_strongly_2015, babin_development_2013, babin_development_2014, medders_development_2014, reddy_accuracy_2016, paesani_getting_2016, chen_ab_2017, marsalek_quantum_2017}. This progress has been greatly facilitated by the development of machine learning potentials (MLPs)~\cite{zhang_deep_2018, wang_deepmd-kit_2018, zeng_deepmd-kit_2023, behler_generalized_2007, fan_neuroevolution_2021, fan_gpumd_2022} that act as efficient surrogates for costly quantum mechanical calculations. 
Studies often considered inaccessible
for direct first-principles simulations became possible, such as predicting the phase diagram of water over a wide range of pressures and temperatures~\cite{reinhardt_quantum-mechanical_2021, zhang_phase_2021, sciortino_constraints_2025}, calculating the $\mathrm{p}K_{\mathrm{w}}$ of liquid water~\cite{calegari_andrade_probing_2023}, estimating the nucleation rate of ice when liquid water is cooled below the freezing point~\cite{piaggi_homogeneous_2022}, etc.
Because of the light hydrogen atoms,
the thermodynamic properties of water in the proximity of the freezing point are affected by
nuclear quantum effects (NQEs). MLPs made it possible to take these effects into account with path-integral molecular dynamics (PIMD) simulations that adopt accurate representations of the Feynman paths~\cite{ceriotti_efficient_2010, cheng_ab_2019, reinhardt_quantum-mechanical_2021, bore_realistic_2023,schran_committee_2020}.

However, validation of these approaches is still incomplete, and systematic benchmarks are needed for both the underlying electronic structure theory and its MLP proxy. This is challenging for properties that require long trajectories. For example, different results have been reported for the equilibrium density of water with different MLPs trained to reproduce the same DFT functional~\cite{cheng_ab_2019, reinhardt_quantum-mechanical_2021, chen_thermodynamics_2024, montero_de_hijes_density_2024}. One model trained on revPBE0-D3~\cite{zhang_comment_1998, grimme_consistent_2010} overestimated the liquid density~\cite{montero_de_hijes_density_2024}, another found a density very close to the experiment~\cite{chen_thermodynamics_2024}, and yet another underestimated it~\cite{cheng_ab_2019}. The manner in which NQEs modify the liquid structure of the
revPBE0-D3 models is also not well established, as the structure is found to be enhanced in Refs.~\cite{marsalek_quantum_2017, ruiz_pestana_quest_2018} while the opposite is found in Ref.~\cite{cheng_ab_2019}. SCAN is another popular functional adopted in water studies~\cite{chen_ab_2017}. SCAN-based MLPs have been adopted in thermodynamic studies of water and ice using classical molecular dynamics (MD) simulations at standard~\cite{piaggi_phase_2021} and at elevated pressures~\cite{zhang_phase_2021}.
It is unknown how the predictions reported in these studies would be affected by NQEs. 

\begin{figure*}[ht!]
\centering
\includegraphics[width=2\columnwidth]{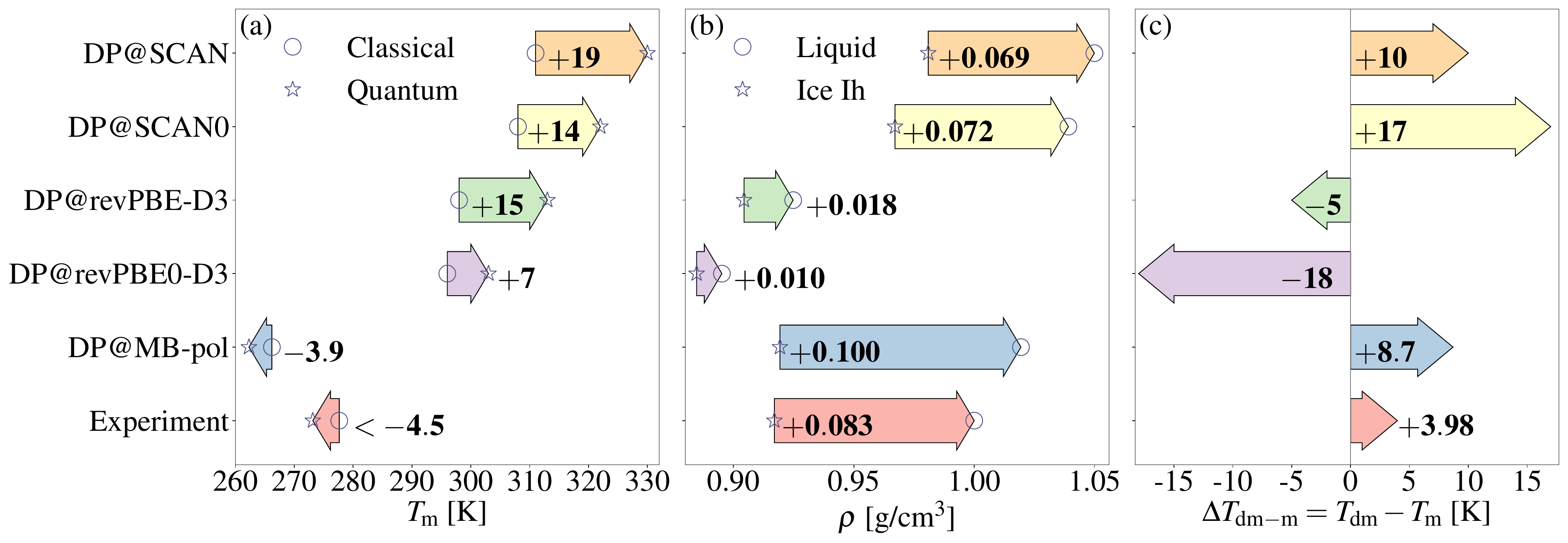}  
\caption{(a) NQEs on the melting temperature of ice at 1 bar, defined as $T_{\mathrm{m}}-T_{\mathrm{m}}^{\mathrm{cl}}$, for DFT-based calculations (this work) and MB-pol (from Ref.~\cite{bore_realistic_2023}). (b) Density discontinuity between water and ice Ih at the calculated melting temperature and 1 bar from quantum simulations. (c) Difference ($\Delta T_{\mathrm{dm}-\mathrm{m}}$) between the temperature of density maximum ($T_{\mathrm{dm}}$) of quantum water and melting temperature ($T_{\mathrm{m}}$) of quantum ice at 1 bar.}
\label{NQEs_Tm_TMD}
\end{figure*}
\begin{table*}[ht!]
\caption{\label{tableTm}Melting Properties of Water at 1 Bar\footnote{The value in parentheses is the statistical uncertainty in the last digit. The values in gray background are quoted from literature and the values in white background are calculated in this work.}}
\begin{ruledtabular}
\begin{tabular}{ccccccccc}
 & $T_{\mathrm{m}}^{\mathrm{cl}}$ [K] & $T_{\mathrm{m}}$
 [K] &  $\Delta T_{\mathrm{m}}^{\mathrm{qu}-\mathrm{cl}}$ [K] &
 $\rho_{\mathrm{liq}}$ [g/cm$^3$] &
 $\rho_{\mathrm{ice}}$ [g/cm$^3$] &
 $\Delta \rho_{\mathrm{liq}-\mathrm{ice}}$ [g/cm$^3$] &
 $T_{\mathrm{dm}}$ [K] &
 $\Delta T_{\mathrm{dm-m}}$ [K] \\
\hline
Experiment~\cite{ceriotti_nuclear_2016}  & \cellcolor{lightgray} $>$277.64 \footnote{Although classical water does not exist in the real world, we can infer that classical H$_2$O would have a melting temperature higher than 277.64 K, given that the melting temperature of T$_2$O and D$_2$O are 277.64 K and 276.97 K, respectively.} &\cellcolor{lightgray} 273.15 &\cellcolor{lightgray} $<-$4.49 &\cellcolor{lightgray} 0.9998 &\cellcolor{lightgray} 0.917 &\cellcolor{lightgray} 0.083 &\cellcolor{lightgray} 277.13 &\cellcolor{lightgray} $+$3.98 \\
q-TIP4P/F~\cite{habershon_competing_2009, ramirez_quantum_2010} &\cellcolor{lightgray} 259 (1) &\cellcolor{lightgray} 251 (1) &\cellcolor{lightgray} $-$8 &\cellcolor{lightgray} 1.003 &\cellcolor{lightgray} 0.921 &\cellcolor{lightgray} 0.082 &\cellcolor{lightgray} 279 (2) &\cellcolor{lightgray} $+$28 \\
DP@MB-pol\footnote{The data for MB-pol are obtained from Ref.~\cite{bore_quantum_2023}.} &\cellcolor{lightgray} 266.2 &\cellcolor{lightgray} 262.3 &\cellcolor{lightgray} $-$3.9 & 1.019 & 0.919 & 0.100 & 271.0 & $+$8.7 \\
DP@revPBE-D3 & 298 (1) & 313 (1) & $+$15 & 0.924 & 0.906 & 0.018 & 308 & $-$5 \\
DP@revPBE0-D3 & 296 (1) & 303 (1) & $+$7 & 0.895 & 0.885 & 0.010 & 285 & $-$18\\
BPNN@revPBE0-D3~\cite{cheng_ab_2019}\footnote{BPNN represents the Behler-Parrinello Neural Network~\cite{behler_generalized_2007}.} &\cellcolor{lightgray} 275 (2) &\cellcolor{lightgray} 267 (2) &\cellcolor{lightgray} $-$8 &\cellcolor{lightgray} 0.94 &\cellcolor{lightgray} 0.89 &\cellcolor{lightgray} 0.05 
&\cellcolor{lightgray} $\approx 277$\footnote{Ref.~\cite{cheng_ab_2019} does not explicitly report the value of $T_{\mathrm{dm}}$. Instead, it states that ``the temperature of density maximum for liquid water matches the experimental value of 3.98 $^\circ$C". Accordingly, we estimate the $T_{\mathrm{dm}}$ reported in this work as 277 K.} 
&\cellcolor{lightgray} $+$10
\\
NEP@revPBE0-D3~\cite{chen_thermodynamics_2024}\footnote{NEP represents the Neuroevolution Potential~\cite{fan_efficient_2017, fan_improving_2022}. Ref.~\cite{chen_thermodynamics_2024} does not study NQEs and all results reported here are from classical simulations.} &\cellcolor{lightgray} 290.5 (5) & & &\cellcolor{lightgray} 1.00 &\cellcolor{lightgray} 0.91 &\cellcolor{lightgray} 0.09 &\cellcolor{lightgray} 292.2 (4) &\cellcolor{lightgray} $+$1.7\\
KbP@revPBE0-D3~\cite{montero_de_hijes_density_2024}\footnote{KbP represents the kernel-based potential~\cite{montero_de_hijes_comparing_2024}. Ref.~\cite{montero_de_hijes_density_2024} does not study NQEs and all results reported here are from classical simulations.} &\cellcolor{lightgray} 246 & & &\cellcolor{lightgray} 1.073 &\cellcolor{lightgray} 0.905 &\cellcolor{lightgray} 0.168 &\cellcolor{lightgray} 220 &\cellcolor{lightgray} $-$26\\
DP@SCAN & 311 (1) & 330 (1) & $+$19 & 1.050 & 0.981 & 0.069 & 340 & $+$10 \\
DP@SCAN0 & 308 (1) & 322 (1) & $+$14 & 1.039 & 0.967 & 0.072 & 339 & $+$17 \\
DP@SCAN (coexistence)\footnote{The coexistence simulation uses 32 beads. The $T_{\mathrm{m}}$ agrees with the $T_{\mathrm{m}}$ from the TI method, with a small discrepancy due to unconverged number of beads, evidenced by $T_{\mathrm{m}}= 328 \pm 1$ K when reducing the beads in the TI calculation. } & & 324 (3) & \\
\end{tabular}
\end{ruledtabular}
\end{table*}

In this work, we provide a systematic benchmark for first-principles-based MLPs for water at standard pressure and introduce a criterion for comparing models and experimental results suggested by the principle of corresponding states. We study how NQEs modify the following properties: (1) the melting temperature of ice; (2) the density discontinuity between water and ice at melting; (3) the temperature of maximum liquid density and its location relative to the melting point. Finally, we compare the liquid models and experiment at effective ``room temperatures", defined to be 25 K above the respective melting points. We consider Deep Potential (DP) models derived from four DFT approximations and the MB-pol method~\cite{babin_development_2013, babin_development_2014, medders_development_2014, reddy_accuracy_2016, paesani_getting_2016}. In the DFT case, we examine revPBE-D3~\cite{zhang_comment_1998, grimme_consistent_2010}, revPBE0-D3~\cite{zhang_comment_1998, grimme_consistent_2010,adamo_toward_1999}, SCAN~\cite{sun_strongly_2015}, and SCAN0~\cite{hui_scan-based_2016}. DP@SCAN indicates the DP model trained on SCAN data, with similar notation for the other models (e.g., DP@SCAN0, BPNN@revPBE0-D3). 
The main findings reported in FIG. \ref{NQEs_Tm_TMD} and TABLE \ref{tableTm} show that DP@MB-pol is closer to experiment for all three properties, while all DFT-based models predict that NQEs raise the melting temperature, contrary to the experimental findings that light-water ice melts at lower temperatures than its heavier isotopes. All DFT-based models underestimate the density discontinuity at melting, but this effect is quite severe with DP@revPBE-D3 and DP@revPBE0-D3. DP@SCAN and DP@SCAN0 correctly predict a density maximum for the liquid above the melting point, but DP@revPBE-D3 and DP@revPBE0-D3 place it below melting. However, at the corresponding ``room temperatures" of 25 K above melting, the liquid structures of all the models show similar features with minor but revealing differences relative to the experiment.      

We use the DP framework~\cite{han_deep_2018, zhang_deep_2018, zhang_end, wang_deepmd-kit_2018, zeng_deepmd-kit_2023} as a surrogate of the first-principles methods. 
revPBE0-D3 and SCAN0 include 25\% and 10\%, respectively, of exact exchange, as recommended in previous studies~\cite{marsalek_quantum_2017,zhang_modeling_2021}. We use an active learning protocol~\cite{zhang_active_2019} to train DP models based on the four DFT functionals using DP-GEN~\cite{zhang_dp-gen_2020} and DeePMD-kit~\cite{wang_deepmd-kit_2018, zeng_deepmd-kit_2023}. This procedure generates the training dataset by exploring the configuration space and training the MLP model, iteratively. In the exploration stage, we use classical MD and quantum PIMD simulations to sample
thermodynamic conditions above and below the melting point at the standard pressure for ice Ih and liquid water. The explored temperature domain ranges
from 270 to 350 K for water and from 150 to 350 K for ice. The active learning protocol builds comprehensive training datasets that include representative configurations of water and ice, whose DFT energies and forces serve as labels for the MLP models. Details, including
training and testing errors,
can be found in Subsection~\ref{prb_MLPs} of the accompanying paper. For the MLP surrogate of MB-pol, we use the DP model trained in Ref.~\cite{bore_realistic_2023} to describe water's phase diagram.

We use LAMMPS~\cite{plimpton_fast_1995, thompson_lammps_2022} to run MD and PIMD simulations in the $NpT$ ensemble at 1 bar, driven by the five DP models introduced above. We adopt a 0.5 fs timestep. PIMD simulations are performed with the ``fix pimd/langevin" module of LAMMPS. In classical MD, the temperature is controlled with a Nos{\'e}-Hoover chain thermostat~\cite{martyna_nosehoover_1992} with a damping time of 0.1 ps, and the pressure is controlled with a Martyna-Tobias-Klein barostat~\cite{martyna_constant_1994} with a damping time of 0.5 ps. In PIMD, the temperature is controlled with a local path integral Langevin equation (PILE\_L) thermostat~\cite{ceriotti_efficient_2010} with a damping time of 0.1 ps, and the pressure is controlled with the Bussi-Zykova-Parrinello barostat~\cite{bussi_isothermal-isobaric_2009} with a damping time of 0.5 ps. We simulate a box of 432 water molecules unless otherwise specified.

In the following, we use the superscript ``cl" to indicate observables calculated with classical MD, and no superscript for PIMD (we use ``qu" for PIMD when required to prevent any ambiguity). We perform thermodynamic integration (TI) following Ref.~\cite{zhang_phase_2021} to calculate the classical chemical potentials of ice and water, $\mu_{\mathrm{ice}}^{\mathrm{cl}}(T)$ and $\mu_{\mathrm{liq}}^{\mathrm{cl}}(T)$, and their difference $\Delta \mu_{{\mathrm{ice}- \mathrm{liq}}}^{\mathrm{cl}}(T)=\mu_{\mathrm{ice}}^{\mathrm{cl}}(T)-\mu_{\mathrm{liq}}^{\mathrm{cl}}(T)$, using the DPTI software~\cite{noauthor_httpsgithubcomdeepmodelingdpti_2024}.
The classical melting temperature ($T^{\mathrm{cl}}_{\mathrm{m}}$) is obtained from the condition $\Delta \mu_{{\mathrm{ice}- \mathrm{liq}}}^{\mathrm{cl}}(T_{\mathrm{m}}^{\mathrm{cl}})=0$. 
Quantum mechanics is converted into classical mechanics by rescaling the Planck constant $\hbar$ with a dimensionless parameter $y$ that continuously switches from 1 to 0. Thus, we can use thermodynamic integration to calculate
the quantum correction to $\mu_{{\alpha}}^{\mathrm{cl}}(T)$ for each phase $\alpha$
~\cite{habershon_competing_2009, cheng_ab_2019, bore_realistic_2023}:
\begin{equation}\label{massti_y_onephase}
\Delta \mu^{\mathrm{qu}- \mathrm{cl}}_{\alpha}(T)=\mu_{\alpha}(T)-\mu^{ \mathrm{cl}}_{\alpha}(T)=\int_{0}^{1}g_{\alpha}(y)\mathrm{d}y,
\end{equation}
with $g_{\alpha}(y)$ defined by
\begin{equation}\label{g_of_y}
g_{\alpha}(y)=2\frac{\braket{K_{\mathrm{CV}, \alpha} \left( \frac{m}{y^2\hbar^2} \right)} - \frac{3Nk_{\mathrm{B}}T}{2N_{\mathrm{H_2O}}}}{y}.
\end{equation}
Here, $K_{\mathrm{CV}, \alpha}$ is the centroid-virial estimator of the quantum kinetic energy per molecule in PIMD for phase $\alpha$. Planck's constant appears in the Feynman paths in combination with the mass $m$
as $\frac{m}{\hbar^2}$; thus, rescaling Planck's constant by $y$ is equivalent to rescaling $m$ by $\frac{1}{y^2}$. The integral over $y$ in
Eq~\eqref{massti_y_onephase} is calculated with the trapezoidal rule on a 7-point grid, including $y=0.0,0.1, 0.2, 0.3, 0.4, 0.6, 1.0$. For each value of $y>0$, we performed PIMD simulations for ice and water in a periodic box with 128 molecules using an increasing number of beads as the quantum limit ($y = 1$) was approached. We found that 8, 16, 32, 64, 64, and 64 beads, respectively, at the 6 grid points from $y=0.1$ to $y=1$ were sufficient for well-converged results. Then, we calculate 
$\Delta \mu_{\mathrm{ice} - \mathrm{liq}}^{\mathrm{qu}}(T)=\Delta \mu^{\mathrm{qu}- \mathrm{cl}}_{\mathrm{ice}}(T)-\Delta \mu^{\mathrm{qu}- \mathrm{cl}}_{\mathrm{liq}}(T)+\Delta \mu_{{\mathrm{ice}- \mathrm{liq}}}^{\mathrm{cl}}(T)$
and estimate $T_{\mathrm{m}}$ from
$\Delta \mu_{\mathrm{ice} - \mathrm{liq}}^{\mathrm{qu}}(T_{\mathrm{m}}) = 0$. In the case of DP@SCAN, we confirm the TI result for $T_{\mathrm{m}}$ with an independent PIMD simulation of direct ice-liquid co-existence for a system of 576 molecules, following Ref.~\cite{piaggi_phase_2021}. Due to the prohibitive cost of quantum coexistence simulations, we used 32 beads instead of the 64 used for TI when $y$ is close to $y=1$. The small difference in $T_{\mathrm{m}}$ calculated with TI and direct co-existence should be attributed to the different number of beads in the two calculations. 

\begin{figure}[ht!]    
\centering
\includegraphics[width=1.0\linewidth]{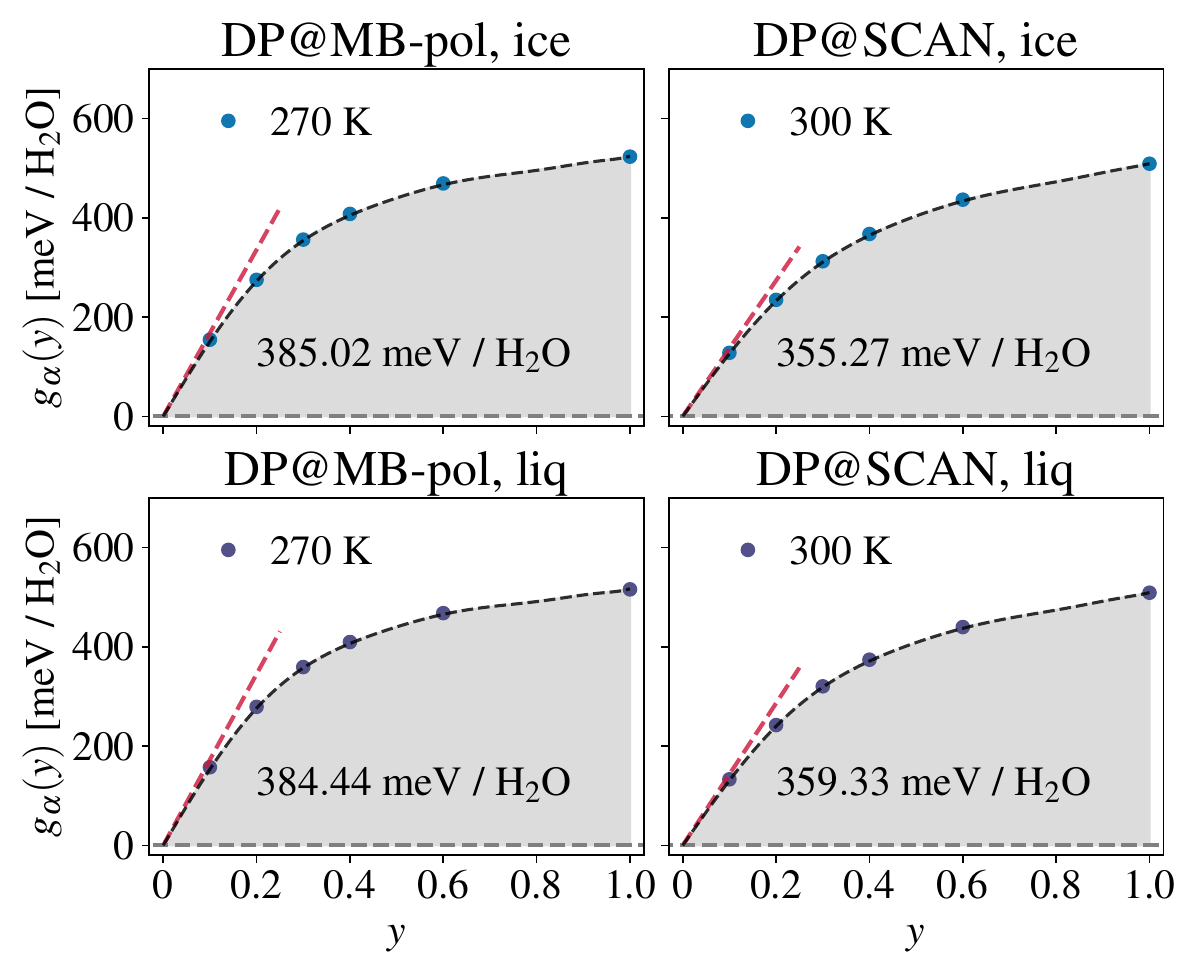}
\caption{The $g(y)$ values for ice and water calculated by DP@MB-pol and DP@SCAN. The dots represent $y$ values at which PIMD are run. The red dashed lines show the slopes $\frac{\mathrm{d}g}{\mathrm{d}y}|_{y=0}$ predicted by an expansion of $\Delta\mu^{\mathrm{qu}- \mathrm{cl}}_{\alpha}(T)$ up to $\hbar^2$ using the perturbation theory. The black dashed lines are polynomial fitting of $g(y)$ using odd orders of $y$ up to $y^{13}$, which corresponds to an expansion of $\Delta\mu^{\mathrm{qu}- \mathrm{cl}}_{\alpha}(T)$ in even orders of $\hbar$ up to $\hbar^{14}$. The shaded area yields the quantum correction to the chemical potential $\Delta\mu^{\mathrm{qu}- \mathrm{cl}}_{\alpha}(T)$ of a single phase $\alpha$.
}
\label{g_y_MB_pol_scan}
\end{figure}

It is instructive to study the variation with $y$ of the integrand in Eq~\eqref{massti_y_onephase}: this is done separately for the correction to the chemical potential of the solid and liquid phases in FIG.~\ref{g_y_MB_pol_scan} in the case of DP@MB-pol and DP@SCAN, which are, respectively, the model that more closely approximates the experimental $T_{\mathrm{m}}$ and the one that deviates more from it. From FIG.~\ref{g_y_MB_pol_scan} we obtain $\Delta \mu^{\mathrm{qu}- \mathrm{cl}}_{\mathrm{ice}}=385.20$ meV / H$_2$O and $\Delta \mu^{\mathrm{qu}- \mathrm{cl}}_{\mathrm{liq}}=384.44$ meV / H$_2$O, respectively, for DP@MB-pol and $\Delta \mu^{\mathrm{qu}- \mathrm{cl}}_{\mathrm{ice}}=355.27$ meV / H$_2$O and $\Delta \mu^{\mathrm{qu}- \mathrm{cl}}_{\mathrm{liq}}=359.33$ meV / H$_2$O, respectively, for DP@SCAN. The quantum correction to the chemical potential is of the same order as the H-bond energies and therefore is not small. The quantum correction to the thermodynamic free energy can be calculated with a perturbative expansion in powers of $\hbar$~\cite{wigner_quantum_1932, uhlenbeck_equation_1932, kirkwood_quantum_1933, landau_statistical_1969_33}. When the statistics of identical particles can be ignored, as is the case here, the expansion contains only even powers of $\hbar$. Thus, $g(y)$, the derivative of the chemical potential with respect to $y$ will contain only odd powers of $y$. The fittings of the data with odd polynomials of the 13th degree in $y$ are shown by the dashed lines in FIG.~\ref{g_y_MB_pol_scan}. The tangent to $g(y)$ when $y=0$ is associated with the lowest-order term in the perturbative expansion. It is easily calculated from classical MD trajectories, as detailed in the accompanying paper, providing an independent test of the accuracy of the PIMD simulations. The melting point is determined by the relative stability of ice and water. The corresponding quantum correction, $\Delta \mu_{\mathrm{ice} - \mathrm{liq}}^{\mathrm{qu}}(T)-\Delta \mu_{\mathrm{ice} - \mathrm{liq}}^{\mathrm{cl}}(T)$, is 2 to 3 orders of magnitude smaller than the quantum correction to the chemical potential of ice or water. Its prediction from first-principles quantum theory is very challenging. $\Delta g_{\mathrm{ice} - \mathrm{liq}}(y)=g_{\mathrm{ice}}(y)-g_{\mathrm{liq}}(y)$ for the five DP models are reported in FIG.~\ref{g_y_funcs} of the accompanying paper. They show a qualitatively similar behavior, but the net results depend on the balance of negative and positive contributions. Of the five models, only DP@MB-pol correctly predicts $\Delta \mu_{\mathrm{ice} - \mathrm{liq}}^{\mathrm{qu}}(T)>\Delta \mu_{\mathrm{ice} - \mathrm{liq}}^{\mathrm{cl}}(T)$.   

Melting properties are summarized in FIG. \ref{NQEs_Tm_TMD} and TABLE \ref{tableTm}, where we also list in gray cells previously reported results in the literature. For the melting temperatures of DP@MB-pol we report the values of Ref.~\cite{bore_realistic_2023}, since we obtain the same results within the error bars of our calculations. The equilibrium density and the temperature at which the liquid has the maximum density ($T_{\mathrm{dm}}$) are obtained from isobars calculated with $NpT$ PIMD for ice and water at different temperatures. These simulations used 32 beads for the Feynman paths and a periodic box with 432 H$_2$O molecules, which allowed accurate calculation of the isobars.  
$\Delta T_{\mathrm{m}}=T_{\mathrm{m}}-T_{\mathrm{m}}^{\mathrm{cl}}$ is the quantum correction to the melting temperature . $\Delta \rho_{\mathrm{liq}-\mathrm{ice}}=\rho_{\mathrm{liq}}-\rho_{\mathrm{ice}}$ is the density discontinuity at melting. 
$\Delta T_{\mathrm{dm}-\mathrm{m}}=T_{\mathrm{dm}}-T_{\mathrm{m}}$ gives the temperature difference between density maximum and melting.
Experimentally, the melting temperature of tritiated water ($\mathrm{T}_2\mathrm{O}$) is higher than that of regular (light) water ($\mathrm{H}_2\mathrm{O}$) by 4.49 K. Since replacing $\mathrm{H}$ with heavier $\mathrm{T}$ makes the water more classical, we infer that $\Delta T_{\mathrm{m}}$ should be smaller than $-4.49$ K. As reported in FIG.~\ref{NQEs_Tm_TMD} (a), all DFT-based DP models instead predict $\Delta T_{\mathrm{m}}>0$ ranging from 7 K (DP@revPBE0-D3) to 19 K (DP@SCAN). The only model with the correct sign of $\Delta T_{\mathrm{m}}$ is DP@MB-pol. Although the quantum correction to the free energy of water or ice is not small, on the order of 3000 K in temperature units, the quantum correction to $T_{\mathrm{m}}$ is more than two orders of magnitude smaller and getting even its sign right is challenging for first-principles quantum mechanical methods. NQEs raise the classical melting point by $\sim$20 K in the case of DP@revPBE0-D3 to $\sim$40 K in the case of DP@SCAN, while it reduces it by $\sim$10 K in the case of DP@MB-pol. Including NQEs, the predicted
$T_{\mathrm{m}}$ deviates more from the experiment with all models. 
As shown in FIG. \ref{NQEs_Tm_TMD} (b), DP@SCAN, DP@SCAN0, and DP@MB-pol predict a density discontinuity in semiquantitative agreement with experiment, but DP@revPBE-D3 and DP@revPBE0-D3 significantly underestimate it. DP@MB-pol, DP@SCAN0, and DP@SCAN overestimate the density of liquid water by 2\%, 4\%, and 5\%, respectively, while DP@revPBE-D3 and DP@revPBE0-D3 underestimate it by 8\% and 10\%, respectively. 
As shown in FIG. \ref{NQEs_Tm_TMD} (c), DP@MB-pol, DP@SCAN, and DP@SCAN0 correctly predict the sign of $\Delta T_{\mathrm{dm}-\mathrm{m}}$, but overestimate its magnitude. In contrast, DP@revPBE-D3 and DP@revPBE0-D3 incorrectly place the maximum liquid density below the melting point.

Additionally, TABLE~\ref{tableTm} reports, for comparison, the predictions of the empirical potential q-TIP4P/F ~\cite{habershon_competing_2009, ramirez_quantum_2010}. As noted in Refs.~\cite{habershon_competing_2009, ramirez_quantum_2010}, q-TIP4P/F yields ice and water densities in good agreement with experiment, but underestimates $T_{\mathrm{m}}$ by approximately 20 K and severely overestimates $\Delta T_{\mathrm{dm}-\mathrm{m}}$. 



\begin{figure}[ht!]    
\centering
\includegraphics[width=1.0\linewidth]{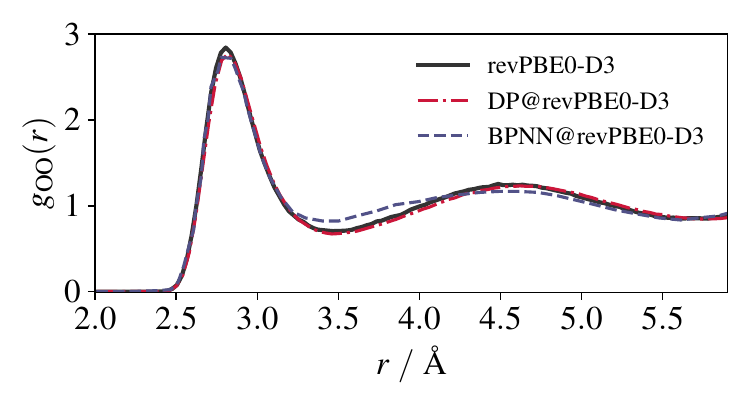}
\caption{$g_{\mathrm{OO}}(r)$ of classical water. The $g_{\mathrm{OO}}(r)$ for revPBE0-D3 is calculated with an AIMD simulation which spans 50 ps. Both DPMD and AIMD simulations are run in the $NpT$ ensemble at 300 K and 1 bar. The $g_{\mathrm{OO}}(r)$ for BPNN@revPBE0-D3 is taken from Ref.~\cite{cheng_ab_2019}.
}
\label{rdfs_compare}
\end{figure}

TABLE~\ref{tableTm} also reports the results of four different MLP models trained on DFT data based on the revPBE0-D3 functional showing a large spread in the model predictions. These differences fall outside the error margins of the MLPs and can arise from model inaccuracy, network architecture, or model training. Due to computational cost some predictions, such as those on the melting temperature and the NQEs correction on it, can only be made with MLP models. Direct comparison with DFT is feasible for the liquid density. As shown in the accompanying paper, the quantum correction on the density is small, and we used classical simulations for this test. Although all revPBE0-D3 models in TABLE \ref{tableTm} give similar predictions for the density of ice, the prediction for the density of the liquid at melting has a wide spread ranging from 0.895 $\mathrm{g/cm^3}$ (DP@revPBE0-D3) to 1.073 $\mathrm{g/cm^3}$ (KbP@revPBE0-D3). We further restrict our attention to DP@revPBE0-D3 and BPNN@revPBE0-D3 because these two models used DFT training data obtained with the CP2K code~\cite{hutter_cp2k_2014} with the settings suggested in Ref.~\cite{monserrat_liquid_2020}, ruling out a possible role of the scheme for electronic structure calculation. We can also rule out a possible effect of the neural network architectures, because a BPNN model trained on our dataset gave the same results as our DP model within statistical errors. This suggests that the difference should come from the training datasets. To validate MLP models against DFT, we ran a 50 ps long \textit{ab initio} molecular dynamics (AIMD) trajectory in the $NpT$
ensemble using CP2K at 300 K and 1 bar with a periodically repeated box containing 64 H$_2$O molecules using the settings of Ref.~\cite{monserrat_liquid_2020}. The O-O radial distribution functions (RDFs) extracted from AIMD, DP@revPBE0-D3, and BPNN@revPBE0-D3 trajectories at the same temperature and pressure
are plotted in FIG.~\ref{rdfs_compare}. Whereas the RDF of DP@revPBE0-D3 matches closely AIMD, the RDF of BPNN@revPBE0-D3 from Ref.~\cite{cheng_ab_2019} shows more radial density in the interstitial region between the first two coordination shells, consistent with a higher liquid density than AIMD or DP@revPBE0-D3. The corresponding differences in the calculated density discontinuity between water and ice at 300 K and 1 bar are relatively modest, 0.04, 0.03, and 0.06 g / cm$^3$ within AIMD, DP@revPBE0-D3 and BPNN@revPBE0-D3. 

However, the differences between the two models based on revPBE0-D3 become significant when the melting temperature and the quantum correction to it are considered. The BPNN@revPBE0-D3 model of Ref.~\cite{cheng_ab_2019} gives
a classical melting temperature $T_{\mathrm{m}}^{\mathrm{cl}}=275$ K, which is lowered by 8 K, when NQEs are taken into account. In contrast, our DP@revPBE0-D3 gives a classical melting temperature $T_{\mathrm{m}}^{\mathrm{cl}}=296$ K, which increases by 7 K when NQEs are included. As a consequence, the differences in the calculated $\Delta \rho_{\mathrm{liq}-\mathrm{ice}}$ and $\Delta T_{\mathrm{dm}-\mathrm{m}}$ 
are enhanced: using BPNN@revPBE0-D3, Ref.~\cite{cheng_ab_2019} reports
$\Delta \rho_{\mathrm{liq}-\mathrm{ice}}=0.06$ g/cm$^3$ and $\Delta T_{\mathrm{dm}-\mathrm{m}}=$ +10 K, while using DP@revPBE0-D3 we find $\Delta \rho_{\mathrm{liq}-\mathrm{ice}}=0.01$ g/cm$^3$ and $\Delta T_{\mathrm{dm}-\mathrm{m}}=-18$ K, respectively.   

The differences between the melting properties predicted by BPNN@revPBE0-D3 and DP@revPBE0-D3 should be attributed almost entirely to the different training datasets. The dataset used by
Ref.~\cite{cheng_ab_2019} includes 1000 classical and 593 quantum water configurations, spanning a wide enthalpy range with a limited number of points near the equilibrium enthalpy of water at 300 K, which limits the accuracy of the model in the vicinity of the melting point. In addition, the dataset did not include ice configurations, which should
play an important role since melting depends on a delicate balance between ice and water. In contrast, the dataset we used to train DP@revPBE0-D3 includes 179 classical water, 749 quantum water, 3 classical ice, and 847 quantum ice configurations. A BPNN trained on our dataset predicted melting properties very close to those of DP@revPBE03-D3, a positive quantum correction to $T_{\mathrm{m}}$, a too small $\Delta \rho_{\mathrm{liq}-\mathrm{ice}}$, and a negative $\Delta T_{\mathrm{dm}-\mathrm{m}}$. These findings suggest that extreme care should be taken when constructing the training dataset, particularly when studying delicate effects like the quantum corrections to the melting properties. In view of the significant change in density when the ice melts, we recommend that $NpT$ simulations be used to collect configurations for DFT training data. For example, the NEP@revPBE0-D3 model in Ref.~\cite{chen_thermodynamics_2024} used a training set geared to simulations in the $NVT$ ensemble. This could be a reason for the overestimated liquid density (1.00 g / cm$^3$) compared with AIMD (0.92 g / cm$^3$).
\begin{figure}[ht!]    
\centering
\includegraphics[width=1.0\linewidth]{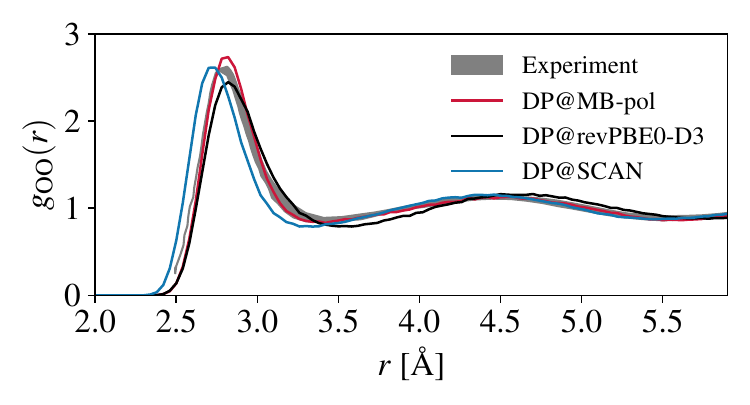}
\caption{$g_{\mathrm{OO}}(r)$ of quantum water at $T_{\mathrm{m}}+25$ K and 1 bar. The experimental $g_{\mathrm{OO}}(r)$ is taken from Ref.~\cite{skinner_benchmark_2013}, with the shaded area indicating the experimental uncertainty. The $g_{\mathrm{OO}}(r)$ curves for DP models are calculated with PIMD simulations in the $NpT$ ensemble.}
\label{rdfs_exp_model}
\end{figure}


Usually, experimental data on the water structure are reported at atmospheric pressure and room temperature, which is conventionally taken to be 25 K above $T_{\mathrm{m}}$. Having determined the melting properties of five MLP models based on first-principles quantum mechanical theory, we can consistently locate the corresponding states of the models relative to their melting point, which should provide a more accurate way to compare these models to the experiment than to make the comparison at the absolute experimental temperature. In FIG.~\ref{rdfs_exp_model} the $g_{\mathrm{OO}}(r)$ of three representative MLP models, DP@MB-pol, DP@revPBE0-D3, and DP@SCAN are compared in this way with the experimental result of Skinner et al.~\cite{skinner_benchmark_2013}. 
Interestingly, the three models compare almost equally well with the experiment, suggesting structural similarity at the corresponding thermodynamic states.
In a closer look, the figure reveals instructive differences. DP@MB-pol deviates from the experiment less than the two other models. DP@SCAN has the first peak displaced to shorter distances but otherwise describes well the order at intermediate range, consistent with the overestimate of the H-bond strength and the enhanced liquid density attributed to the SCAN functional approximation~\cite{chen_ab_2017}. DP@revPBE0-D3 has the first and second peaks displaced at larger distances, consistent with the underestimation of the liquid density associated with this model.   We did not report the RDFs of DP@SCAN0 and DP@revPBE-D3 in FIG.~\ref{rdfs_exp_model}, as these RDFs are very similar to those of DP@SCAN and DP@revPBE0-D3, respectively, as shown in FIG.~\ref{gOOs-hybrid} of the accompanying paper.

The NQEs on the RDFs of liquid water, calculated at their corresponding states, are compared with the experimental RDFs of H$_2$O and D$_2$O. As shown in FIG.~\ref{gOOs-6} of the accompanying paper, all models correctly reproduce the experimental trend that NQEs slightly soften the first peak of $g_{\mathrm{OO}}(r)$ while introducing negligible changes in the second peak. Although the models underestimate the magnitude of the experimental H$_2$O–D$_2$O difference, particularly in the first interstitial region, their overall agreement with experiment is good. These modifications to the RDFs induced by NQEs further emphasize the need to calculate the classical and quantum structural properties at their corresponding states. By contrast, when the classical and quantum RDFs are compared at the same absolute temperature, the quantum RDF exhibits an artificial enhanced structure in the second peak due to the quantum effect, as shown in FIG.~\ref{gOOs-absoluteT} of the accompanying paper.

We conclude our paper with several remarks. First, we have assessed the capabilities of the MLPs based on DFT and MB-pol to describe the anomalies of water related to the melting of ice. As expected, MB-pol provides the most accurate reproduction of the melting properties. Among the four DFT functionals considered, SCAN offers the best balance between accuracy in predicting these properties and computational cost.
Second, we emphasize the importance of sufficiently diverse datasets for MLPs. MLPs are proxies for the underlying quantum mechanical models and extreme care should be taken when using these potentials for the prediction of a property as delicate as the sign of the isotope effect on the melting temperature of ice.
Finally, accurately modeling the hydrogen bond strength is important for predicting the properties of water. To achieve this goal, better functionals for DFT or accurate quantum chemistry methods beyond DFT are required.

We thank Ruiqi Gao, Han Wang, Jinzhe Zeng, and Linfeng Zhang for useful discussions. This work is supported by the Computational Chemical Sciences Center “Chemistry in Solution and at Interfaces” under Award No. DE-SC0019394 from the U.S. Department of Energy. We acknowledge the computational resources provided by the National Energy Research Scientific Computing Center (NERSC), which is supported by the U.S. Department of Energy (DOE), Office of Science under Contract No. DE-AC0205CH11231 and Princeton Research Computing at Princeton University. P.M.P.~acknowledges funding from the Marie Skłodowska-Curie Cofund Programme of the European Commission project H2020-MSCA-COFUND-2020-101034228-WOLFRAM2. The models, data and input files supporting the calculations can be found at our GitHub repository~\cite{li_httpsgithubcomyi-fanlinqe-ice-tm_nodate}. 

\bibliography{clean}

\end{document}